\documentclass[letter]{aa}
\usepackage{amsmath}
\usepackage{graphicx}
\usepackage{breqn} %para splitear ecuaciones 
\usepackage[mathscr]{euscript} %para tipografía especial
\usepackage{comment}
\usepackage[utf8]{inputenc}

\usepackage{booktabs}
\usepackage{adjustbox}
\usepackage{tablefootnote}

\usepackage{tablefootnote}
\usepackage{caption}
\usepackage{subcaption}
\usepackage{ulem}

\usepackage{multicol}

\usepackage{multirow}

\usepackage[]{hyperref}
\hypersetup{unicode=true, colorlinks=true, linkcolor=[rgb]{0.53, 0.15, 0.34}, citecolor=[rgb]{0.0, 0.27, 0.42}, filecolor=[rgb]{1.0, 0.13, 0.32}, urlcolor=[rgb]{0.53, 0.15, 0.34}}

\usepackage{natbib} %For the bibliography
\bibpunct{(}{)}{;}{a}{}{,} % to follow the A&A style

\usepackage{txfonts}
\graphicspath{{Graphics/}}

\begin{document}

   \title{Microquasar remnants as hidden PeVatrons}
   \author{Leandro Abaroa\inst{1,2}\thanks{leandroabaroa@gmail.com}, 
          Gustavo E. Romero\inst{1,2},
           \and Valentí Bosch-Ramon\inst{3}
           }

   \offprints{Leandro Abaroa}
  \institute{Instituto Argentino de Radioastronom\'ia (CCT La Plata, CONICET; CICPBA; UNLP), C.C.5, (1894) Villa Elisa, Argentina \and Facultad de Ciencias Astron\'omicas y Geof\'{\i}sicas, Universidad Nacional de La Plata, B1900FWA La Plata, Argentina \and Departament de F\'isica Qu\`antica i Astrof\'isica, Institut de Ci\`encies del Cosmos (ICC), Universitat de Barcelona (IEEC-UB), Mart\'i i Franqu\`es 1, E08028 Barcelona, Spain}

   \date{Received / Accepted}

%-------------------------------------------------------

\abstract{The Large High Altitude Air Shower Observatory (LHAASO) has revealed numerous ultrahigh-energy gamma-ray sources of unknown origin. We propose that a fraction of them can be explained by microquasar remnants, i.e., binary systems where mass transfer has ceased and the central engine is quenched. Cosmic rays injected during the active phase of a microquasar may remain confined within its cocoon and subsequently interact with nearby molecular clouds, producing bright gamma-ray emission through \textit{pp} collisions. Remnants of former super-Eddington systems can act as dark PeVatrons, releasing particles up to $\sim$10 PeV that illuminate surrounding clouds %with 
producing gamma rays reaching hundreds of TeV. This scenario provides a natural explanation for several unidentified Galactic LHAASO sources.}

\authorrunning{Abaroa et al.}
\titlerunning{Microquasars remnants as hidden PeVatrons}

\maketitle
\section{Introduction}\label{sec: intro}

The terminal regions of microquasar (MQ) jets are efficient sites of particle acceleration to relativistic energies \citep{Heinz&Sunyaev_2002,2005A&A...429..267B,2008A&A...485..623R,2009A&A...497..325B}. Recent studies show that super-Eddington MQs and binaries can accelerate particles up to $\gtrsim 10$ PeV \citep{Abaroa_etal2024(S26),Peretti2025,Zhang2025arXiv250620193Z}. Energetic arguments further indicate that transrelativistic jets and winds with kinetic power $\gtrsim 10^{39}\,{\rm erg\,s^{-1}}$ are the most likely Galactic PeVatrons \citep{Wang2025ApJ}, strengthening the case for MQs as sources of cosmic rays (CRs) beyond the knee. These predictions are consistent with the ultrahigh-energy (UHE) gamma rays recently detected by the Large High Altitude Air Shower Observatory (LHAASO) 
\citep{LHAASO-Catalog2024ApJS,LHAASO2024a,LHAASO2024b}.

%Among the UHE sources detected by LHAASO, five are associated with MQs.\textbf{MODIFICAR ACÁ: 'LA MAYORÍA SON SUPER EDD...'} Except for Cyg~X-1, all host or are suspected to host compact objects accreting at super-Eddington rates %superaccreting

Several UHE sources detected by LHAASO are associated with MQs. Most of these sources are believed to host compact objects that accrete matter at super-Eddington rates, either permanently or during certain phases of their activity \citep{LHAASO2024a}. This connection suggests that not only active MQs but also their long-lived remnants could contribute to the unidentified LHAASO population.

In this \textit{Letter}, we propose that microquasar remnants (MQRs) can act as hidden PeVatrons. In an MQR, mass transfer from the companion star to the black hole (BH) ceases permanently. This may occur if the system loses angular momentum, widening the orbit from semi-detached to detached \citep{Willcox_etal_2023}; if a massive companion ($M_*>40,M_{\odot}$) collapses directly into a BH without a supernova \citep{Belczynski_etal_2016(Nature)}; or if a strong wind from a supercritical accretion disk expels the stellar envelope \citep{Ohsuga2005,Abaroa_etal_2023,Abaroa&Romero_RevMex_2024}.

Once the central engine shuts down, the only fossil is the non-interacting binary at the center of the cocoon inflated by the jets during the MQ’s lifetime. This structure resembles Fanaroff-Riley II cocoons \citep{Begelman&Cioffi_1989}, but on smaller scales \citep{2009A&A...497..325B, Abaroa_etal2024(S26)}.

\begin{figure}
    \centering
    \includegraphics[width=\columnwidth]{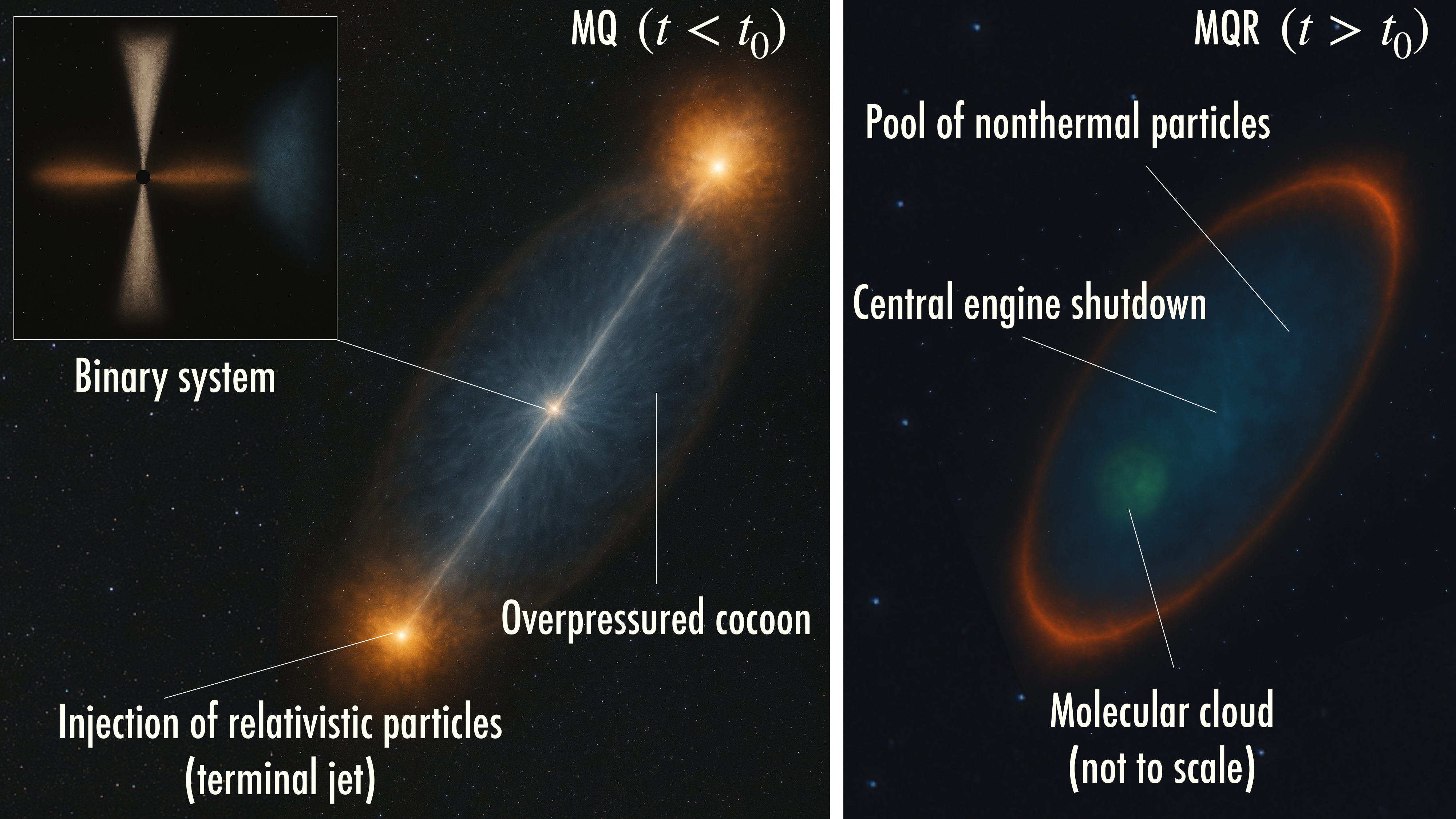}
    \caption{\footnotesize Conceptual scheme, not to scale. \textit{Left panel:} MQ ($t<t_0$). The stellar-mass BH and the star are in the core. The backward shock at the end of the jets accelerates particles to relativistic energies. An overpressured cocoon evolves with the jets and envelops the entire system. \textit{Right panel:} MQR ($t>t_0$). The central engine shuts down, and an MQR forms. CRs distributed within the survivor cocoon interact with a fragment of molecular cloud that enters the MQR, illuminating it.}
    \label{fig: scheme}
\end{figure}

Ultra-relativistic CRs injected throughout the MQ’s active phase can irradiate dense clumps or cloud fragments engulfed by the cocoon, producing gamma rays via $pp$ interactions. A schematic of this scenario is shown in \hyperref[fig: scheme]{Fig.~\ref{fig: scheme}}.

CR diffusion timescales depend on the diffusion coefficient, $\tau = R^2 / D(E)$, where $D(E) \propto E^{\delta}$ ($\delta = 1$ in the Bohm limit), so the highest-energy protons diffuse more rapidly and reach the clouds first. The duration of cloud irradiation and the resulting gamma-ray spectrum depend on several parameters, while escaping CRs can propagate through the interstellar medium (ISM) and interact with more distant molecular clouds, potentially producing multiple, spatially separated gamma-ray sources with spectra set by the diffusion coefficient \citep{1996A&A...309..917A,2005A&A...432..609B}.

In what remains of this \textit{Letter}, we first present a model for MQRs (\hyperref[sect: MQR]{Sect. \ref{sect: MQR}}) and then the CR propagation model (\hyperref[sect: relativistic particles]{Sect. \ref{sect: relativistic particles}}), show the results of CR–cloud interactions (\hyperref[sect: results]{Sect. \ref{sect: results}}), and conclude with a discussion (\hyperref[sect: conclusion]{Sect. \ref{sect: conclusion}}).

\section{The microquasar remnant}\label{sect: MQR}

We assume that the MQ remains active for $t_{0} = 10^5\,{\rm yr}$, injecting power through twin jets into the ISM and carving out a dilute bubble that is subsequently filled with relativistic particles accelerated in the shocks formed at the jet termination regions. This dilute bubble corresponds to the hot cocoon inflated by the bow shocks, which are driven by the jets as they propagate through the ISM. Such jet-inflated bubbles have been extensively studied and observed in both radio galaxies and MQs,  supporting the adopted physical picture \citep{Begelman&Cioffi_1989, 1997MNRAS.286..215K,Heinz&Sunyaev_2002,kino&kawakatu2005, 2009A&A...497..325B, 2010Natur.466..209P, soria2010,Marti_etal_2015,Marti_etal_2017NatCo, Abaroa_etal2024(S26)}. 

Once the injection ceases, the system evolves into an MQR. We adopt a BH mass of $10\,M_{\odot}$ accreting (for $t < t_0$) at super-Eddington rates with an accretion power at the jet launch radius of $L_{\rm acc} \approx 10^{41}\, {\rm erg \,s^{-1}}$. \hyperref[tab: parametros generales]{Table~\ref{tab: parametros generales}} in Appendix \ref{app: table}  lists the main parameters of our model.

\subsection{Jet}\label{sec: jet}

According to the disk–jet coupling hypothesis (e.g., \citealt{1995A&A...293..665F}), the kinetic power of each jet is assumed to be a fraction of the accretion power at the launching radius, $L_{\rm j}=q_{\rm j} L_{\rm acc}$. Typical jet powers of Galactic and extragalactic MQs range from $10^{37}$ to $10^{41}\,{\rm erg\, s^{-1}}$ (e.g., \citealt{Heinz&Sunyaev_2002,2010Natur.466..209P}). 
We adopt $q_{\rm j}=0.1$, yielding $L_{\rm j}=10^{40}\,{\rm erg\, s^{-1}}$ per jet as a representative value supported by observations of powerful systems such as SS 433 and S26 in NGC 7793, whose large-scale radio and X-ray bubbles imply sustained kinetic luminosities of $>10^{40}\,{\rm erg\,s^{-1}}$ over $> 10^5\rm \, yr$. Such powers are naturally explained by super-Eddington accretion onto the compact object. We also assume a moderate value for the jet Lorentz factor, $\gamma_{\rm j}=(1-v_{\rm j}^2/c^2)^{-1/2}=3$, which corresponds to a velocity $v_{\rm j}\approx 0.943\,c$ \citep{Heinz&Sunyaev_2002}.

The jet head propagates through the ISM, and its length $l_{\rm j}$ at a time $t$ $(<t_0)$ is approximated here as $l_{\rm j}=(L_{\rm j}/\rho_{\rm ISM})^{1/5}\,t^{3/5}$ \citep{1997MNRAS.286..215K}, where the typical intercloud ISM density is $\rho_{\rm ISM}=1.7\times10^{-25}\,{\rm g\, cm^{-3}}$ \citep{Spitzer-Book1978}. At $t=t_0$, the jet length is $\approx 2.8\times10^{20}\,{\rm cm}\approx 95\,{\rm pc}$. The longitudinal velocity of the jet head is $v_{\rm l}={\rm d}l_{\rm j}/{\rm d}t=3l_{\rm j}/5t$, which at $t=t_0$ gives $5.4\times10^{7}\,{\rm cm\,s^{-1}}$. An adiabatic reverse shock (RS) develops at the jet termination region, where it accelerates particles that are then advected by the bowshock backflows and subsequently injected into the cocoon. Once accelerated at the RS, relativistic particles are entrained in the downstream flow of the shocked jet material. This post-shock plasma streams backwards along the cocoon, advecting and mixing the relativistic particles throughout its volume, where they can remain confined for long times before diffusing out. 

The RS is a strong, relativistic shock %($v_{\rm sh}\approx  0.5c$) 
capable of efficient diffusive shock acceleration, as inferred in large-scale jet termination regions. Observations of non-thermal lobes in several MQs (e.g., \citealt{2010Natur.466..209P, Alfaro_etal_v4641_2024Nature}) demonstrate that particles can indeed be accelerated at such termination shocks and later advected into the cocoon.%Details of the calculations presented in this section can be found in \cite{Abaroa&Romero_2024(cluster)} and \cite{Abaroa_etal2024(S26)}.

Shortly after mass transfer stops, the BH accretes all matter from the accretion disk. Once the accretion stops, the system can no longer produce jets. %\textbf{The relativistic particles previously accelerated, however, remain trapped within the cocoon, as described below.}

\subsection{Cocoon}\label{sect: cocoon}

The cocoon is a cavity inflated by the jet that remains overpressured and expands into the ISM with an ellipsoidal geometry if the medium is approximately homogeneous. Its semi-major axis equals the jet length during the active MQ phase, $l_{\rm c}(t<t_0)=l_j(t<t_0)$, while the semi-minor axis at any $t$ is $w_{\rm c}\sim l_{\rm c}/3$. This aspect ratio for the cocoon is consistent with analytical and observational results for jet-driven lobes (e.g., \citealt{Begelman_etal_1984_radiogalaxies,1997MNRAS.286..215K, 2009A&A...497..325B, soria2010}). The cocoon volume is taken here to be $V_{\rm c}(t)=4\pi l_{\rm c}(t)^3/27$.

%After $t_0$, when accretion ceases and the jets no longer propagate, the cocoon continues to expand as an adiabatic bubble powered by an impulsive energy injection. Its longitudinal evolution is given by $l_{\rm c}(t>t_0)=(L_{\rm j}\,t_0/\rho_{\rm ISM})^{1/5}\,t^{2/5}$, with a corresponding expansion velocity $v_{\rm c}={\rm d}l_{\rm c}/{\rm d}t$. This adiabatic phase lasts until the onset of the radiative phase $(t_{\rm rad})$. The transition occurs when the cooling time $t_{\rm cool}(t)=3k_{\rm B}T(t)/n_{\rm ISM}\Lambda(T(t))$ becomes equal to $t$ \citep{castor_etal_1977}, where $k_{\rm B}$ is the Boltzmann constant, $\Lambda$ the cooling function, and $T(t)=2E_{\rm int}(t)/3n_{\rm ISM}V_{\rm c}(t)k_{\rm B}$ (with $E_{\rm int}(t)\sim 0.5L_{\rm j}t$). For the parameters of our model, we find $t_{\rm rad}\gg t_0$.

After $t_0$, when accretion ceases and the jets no longer propagate, the cocoon continues to expand as an adiabatic bubble powered by an impulsive energy injection. Its longitudinal evolution is approximated as $l_{\rm c}(t>t_0)=(L_{\rm j}\,t_0/\rho_{\rm ISM})^{1/5}\,t^{2/5}$, with a corresponding expansion velocity $v_{\rm c}={\rm d}l_{\rm c}/{\rm d}t$. This adiabatic phase lasts until the onset of the radiative phase $(t_{\rm rad})$ (see Appendix \ref{app: cocoon survival}).

\section{Production and distribution of cosmic rays}\label{sect: relativistic particles}

A fraction $q_{\rm rel}\approx0.1$ of the jet power is assumed to be converted to relativistic particles: $L_{\rm rel}=q_{\rm rel}\,L_{\rm j}$. We include both the hadronic and leptonic populations, $L_{\rm rel}=L_{\rm p}+L_{\rm e}$. %The total energy of CRs injected is $E_{\rm CR}=L_{\rm rel}\,t_{0}$, so we have $3\times10^{51}\,{\rm erg}$. 
The energy distribution between hadrons and leptons is unknown. We assume a hadron-dominated scenario ($L_{\rm p}=100\,L_{\rm e}$), meaning that  99\%  of the non-thermal power resides in protons while only  1\%  is carried by electrons. This is consistent with shock acceleration efficiencies inferred in supernova remnants and MQ jets (e.g., \citealt{2008A&A...485..623R}), and is also supported by particle-in-cell simulations \citep{caprioli&Spitkovsky2014_PIC}.

The diffusive acceleration rate of the particles is given by $t^{-1}_{\rm{acc}}=\eta_{\rm acc}e\,Z\,c\,B/E$,
where $e$ is the electric charge, $Z$ the atomic number, $c$ the speed of light, $B$ the magnetic field, and $E$ is the energy of the particle. We assume an equipartition fraction of $\sim 0.1$ between the magnetic pressure and thermal pressure to determine the strength of the magnetic field in the post-shock region ($B\sim 10\mu\rm G$). We note that our results remain essentially the same if we vary the value of $B$ in a reasonable parameter range. We also note that this value corresponds to a few times the compressed interstellar value and is well below equipartition with the jet luminosity. This ensures efficient acceleration without requiring unrealistically strong fields. On the other hand, the upstream medium is expected to have a weaker field and a larger diffusion coefficient. Since particle confinement is primarily limited by advection through the downstream region, we use the downstream $B$ to estimate $E_{\rm max}$. 

The acceleration efficiency of the process in the non-relativistic limit can be estimated as $\eta_{\rm acc}\simeq3\beta_{\rm sh}^2/8$ \citep{1983RPPh...46..973D}, where we adopt Bohm diffusion\footnote{This assumption is usually done in the acceleration region at the terminal shocks of the jets (e.g., \citealt{SS433-2018Natur.562...82A,SS433-2024Sci}).}%We note that assuming near-Bohm diffusion represents an upper limit for acceleration efficiency. Mildly relativistic shocks in MQ jets can sustain strong turbulence and partial self-generation of magnetic irregularities, making Bohm diffusion plausible in such environments. This assumption was confirmed, for example, in the supernova remnant Cassiopea A \citep{Cassiopea2006Nature} and in the MQ SS 433 \citep{SS433-2018Natur.562...82A,SS433-2024Sci}.
at the shocks and where $\beta_{\rm sh}=v_{\rm sh}/c$ (with $v_{\rm sh}$ the shock velocity, see \hyperref[tab: parametros generales]{Table~\ref{tab: parametros generales}}). These values lead to $\eta_{\rm acc}\approx0.1$. Escape is advective in the terminal region of the jet $t_{\rm adv}^{-1}=(4\Delta x/v_{\rm j})^{-1}$, and diffusive in the cocoon $t_{\rm diff}^{-1}=(\Delta x^2/D(E))^{-1}$, where $\Delta x$ is the size of the corresponding region. Here, $D(E)$ is the diffusion coefficient, considered as a factor $\zeta$ of the Bohm coefficient: $D(E)=\zeta D_{\rm B}(E) =\zeta E\,c/3\,B\,e$. To make some numerical estimates, we adopt $\zeta=5$ in the interior of the cocoon. This choice reflects the expected turbulence of the shocked jet plasma, which is likely filled with magnetic irregularities that strongly scatter CRs. Such a suppressed diffusion, while not strictly Bohmian, is consistent with conditions inferred in analogous systems such as radio-galaxy lobes and the Galactic Fermi bubbles, where relativistic particles appear confined for extended periods of time (e.g., \citealt{crocker&aharonian2011Fermibubbles}).

For simplicity, we neglect further acceleration mechanisms that might occur within the cocoon, such as Fermi type II or reacceleration due to the interaction between the CRs and the cocoon's shell.
In other words, all acceleration occurs at the RS, and the cocoon merely serves as a reservoir that traps the particles that have already been accelerated. %This prolonged confinement enables significant pp interactions with any embedded clouds, whereas a faster (ISM-like) diffusion would lead to a much shorter residence time and a substantially fainter gamma-ray output.}

Relativistic particles cool via nonthermal processes arising from interactions with radiation, magnetic fields, and matter, and also experience adiabatic losses in the RS and cocoon due to work done during their expansion. The following subsections describe the particle distribution in the cocoon during the MQ and MQR phases and CR propagation through the ISM. From this point onward, we focus on protons, as electrons are expected to cool rapidly via synchrotron losses.  

%\subsection{Particle generation: MQ}

The relativistic particles are accelerated at the reverse shock in the moving head of the jets. We estimate this population solving the time-dependent transport equation in the RS during the lifetime of the MQ ($t<t_0$).

%This equation is: 
%\begin{equation}\label{eq: transport_mq}
%   \frac{{\partial n_{\rm RS}(E,t)}}{\partial t}+ \frac{\partial}{\partial E}\left[\frac{{\rm d}E}{{\rm d}t} n_{\rm RS}(E,t)\right]+\frac{n_{\rm RS}(E,t)}{t_{\rm esc,RS}(E,t)}=Q_{\rm RS}(E,t).
%\end{equation}
The injection function for the source, $Q_{\rm RS}(E,t)=Q_{0}(t)E^{-p}\exp{(-E/E_{\rm max}(t))}$, is a power-law in energy with an exponential cutoff and a spectral index of $p= 2$, which is characteristic of the Fermi first-order acceleration mechanism \citep[see, e.g.,][]{1983RPPh...46..973D}. 
%The normalization constant $Q_0$ is obtained from the available power for the CRs. 
At $t_0$, the particle injection stops $(Q_0=0)$, and the fluid rapidly evacuates the acceleration region by advection $(t_{\rm esc,RS}\equiv t_{\rm adv})$. Thus, the particle distribution is completely suppressed for $t\gtrsim t_0$. We note that, to achieve energies greater than 1 PeV, the source must be active for at least $t_{\rm acc}\sim 3 E_{\rm PeV}\eta^{-1}_{0.1} B^{-1}_{ 10\mu \rm G}$ yr. This means that, if the source is sporadic, it should have a high duty cycle to secure the necessary total power, with active periods lasting several years to reach adequate energies. In Appendix \ref{app: cocoon transport} we detail the equation for particle transport in the cocoon.  

%\subsection{Particle distribution: MQR}

\begin{figure}[ht!]
  \centering
  \begin{minipage}{0.5\textwidth}
    \centering
    \includegraphics[width=9cm, height=4.9cm]
    {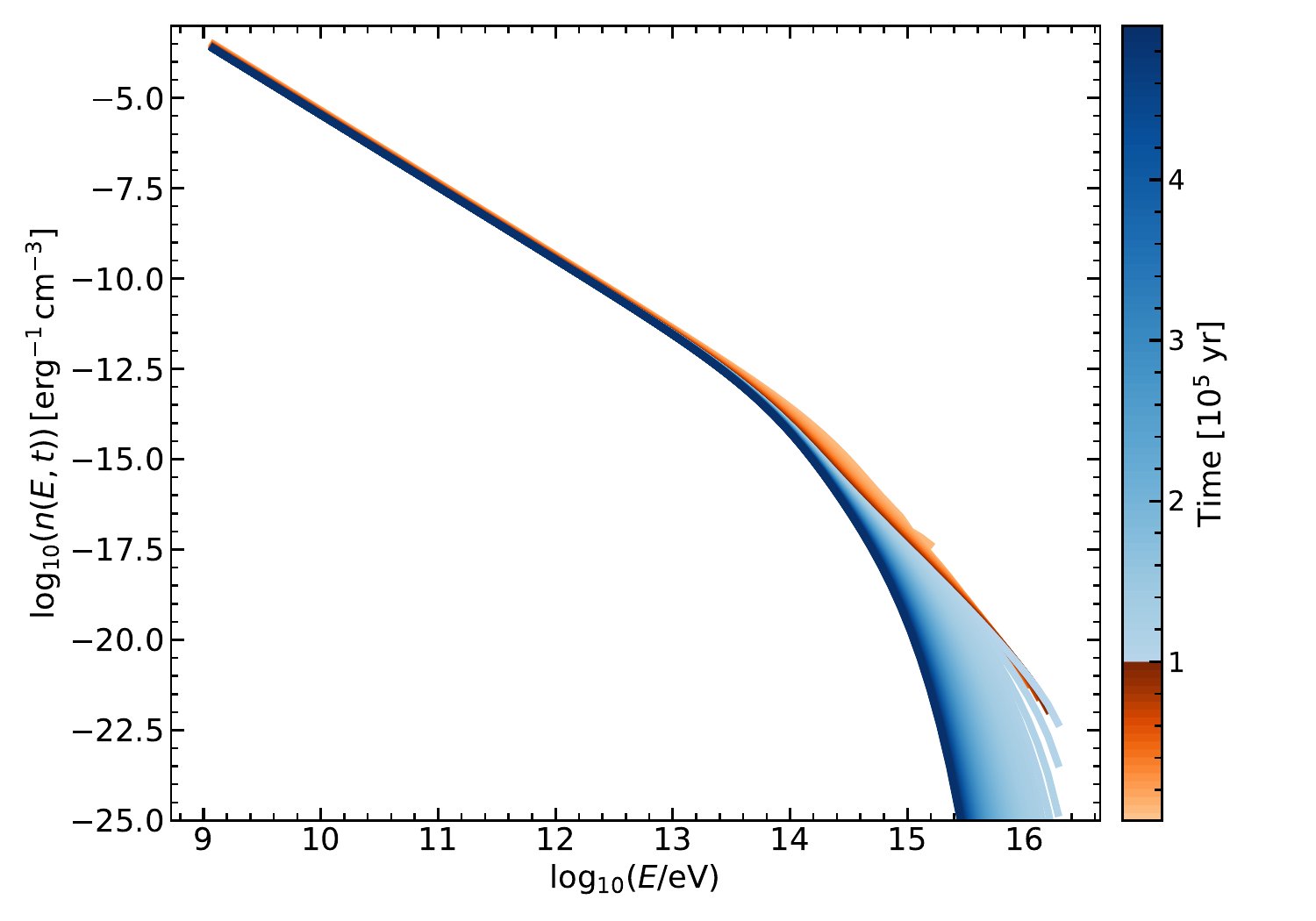}
  \end{minipage}

  \vspace{-0.2cm} % <-- reduce separación vertical

  \begin{minipage}{0.5\textwidth}
    \centering
    \includegraphics[width=8cm, height=4.9cm]{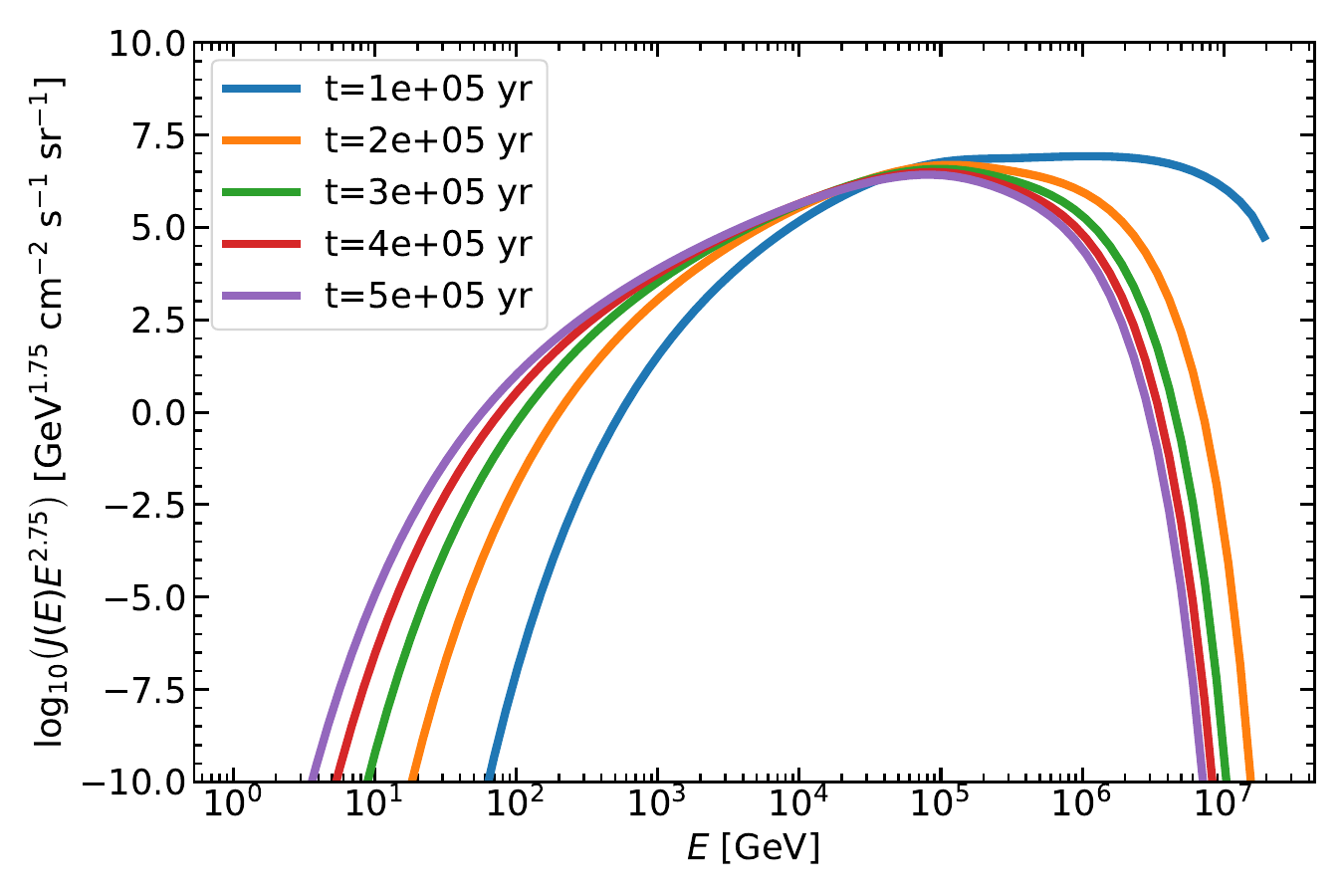}
  \end{minipage}
  \vspace{-0.4cm} % <-- reduce separación vertical

  \caption{\footnotesize Evolution of proton distributions. \textit{Top}: CRs within the cocoon. The color bar distinguishes the MQ ($t<t_0$, orange) and MQR ($t>t_0$, blue) phases. \textit{Bottom}: CR propagation to a distance of 100 pc from the MQR over time, for a source with a continuous injection of particles.}
    \label{fig: proton_distribution}
\end{figure}

\hyperref[fig: proton_distribution]{Figure~\ref{fig: proton_distribution}} (Top) shows the evolution of the proton distribution in the cocoon up to $t=5\times 10^5\,{\rm yr}$. It can be seen that, even hundreds of thousands of years after the MQ is turned off, particles with energies $> 1\,$PeV remain within the cocoon. The maximum energy of the particles depends on time; we find that, under Bohm regime and at $t_0$, $E_{\rm max}\approx 25 \,{\rm PeV}$. %Non-Bohmian diffusion in the cocoon is explored for comparison in the Appendix.}

\begin{comment}
\begin{figure}[ht!]
  \centering
  \begin{minipage}{0.5\textwidth}
    \centering
    \includegraphics[width=9.6cm, height=5.9cm]{graphics/distribution_proton_cocoon.pdf}
  \end{minipage}
  %\vspace{-0.8cm} % <-- reduce separación vertical
  \begin{minipage}{0.5\textwidth}
    \centering
    \includegraphics[width=9cm, height=5.8cm]{graphics/proton_propagation_ism_continuous_fixed_distance.pdf}
  \end{minipage}
%  \vspace{-0.4cm} % <-- reduce separación vertical

  \caption{\footnotesize Evolution of proton distributions. \textit{Top}: CRs within the cocoon. The color bar distinguishes the MQ ($t<t_0$, orange) and MQR ($t>t_0$, blue) phases. \textit{Bottom}: CR propagation to a distance of 100 pc from the MQR over time, for a source with a continuous injection of particles.} %an impulsive source.}
  \label{fig: proton_distribution}
\end{figure}
\end{comment}

\section{Irradiation of clouds}\label{sect: results}

\begin{comment}
\begin{figure}[ht!] \label{fig: seds}
  \centering
  \begin{minipage}{0.5\textwidth}
    \centering
    \includegraphics[width=9.4cm, height=5.7cm]{graphics/sed_cloud_cocoon.pdf}
  \end{minipage}
  \vspace{-0.4cm} % <-- reduce separación vertical
  \begin{minipage}{0.5\textwidth}
    \centering
    \includegraphics[width=9.4cm, height=5.7cm]{graphics/SED_cloud_ism_continuous.pdf}
  \end{minipage}
  %\vspace{-0.1cm} % <-- reduce separación vertical

  \caption{\footnotesize SEDs of the illuminated clouds at different times (see color bar). Leptonic emission is produced by secondary pairs created in the cloud via $pp$ interactions. Thick dashed lines show the sensitivity curves of \textit{Fermi} (10 yr, grey), SWGO (5 yr, light blue), and LHAASO (1 yr, green) for a Galactic MQR at 5 kpc. \textit{Top}: Cloud in the MQR. \textit{Bottom}: Cloud at 100 pc from the MQR (continuous scenario).}
  \label{fig: seds}
\end{figure}
\end{comment}
\begin{figure}[ht!]
  \centering
  \begin{minipage}{0.5\textwidth}
    \centering
    \includegraphics[width=8.4cm, height=4.9cm]{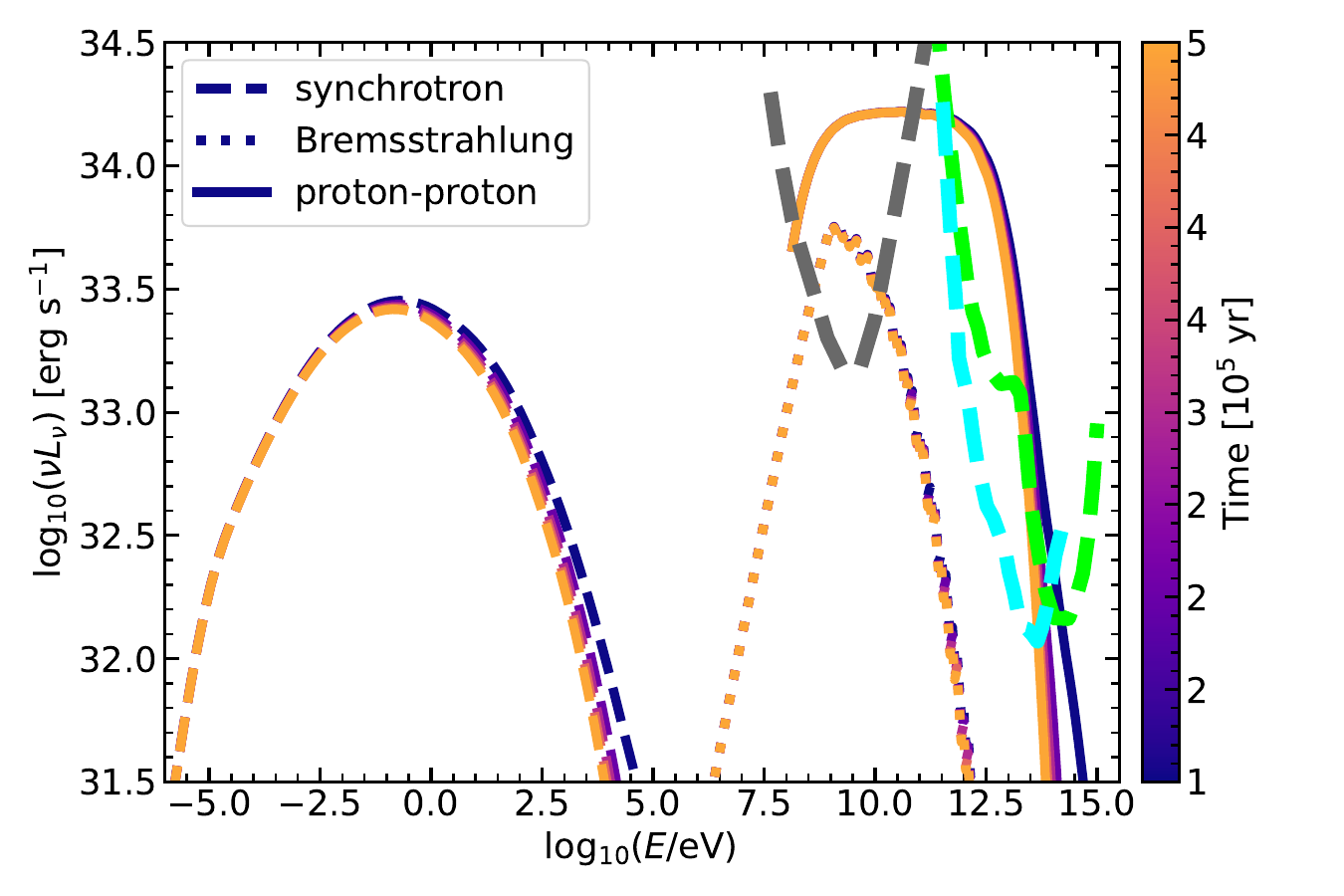}
  \end{minipage}

  \vspace{-0.2cm} % <-- reduce separación vertical

  \begin{minipage}{0.5\textwidth}
    \centering
    \includegraphics[width=8.4cm, height=4.9cm] 
    {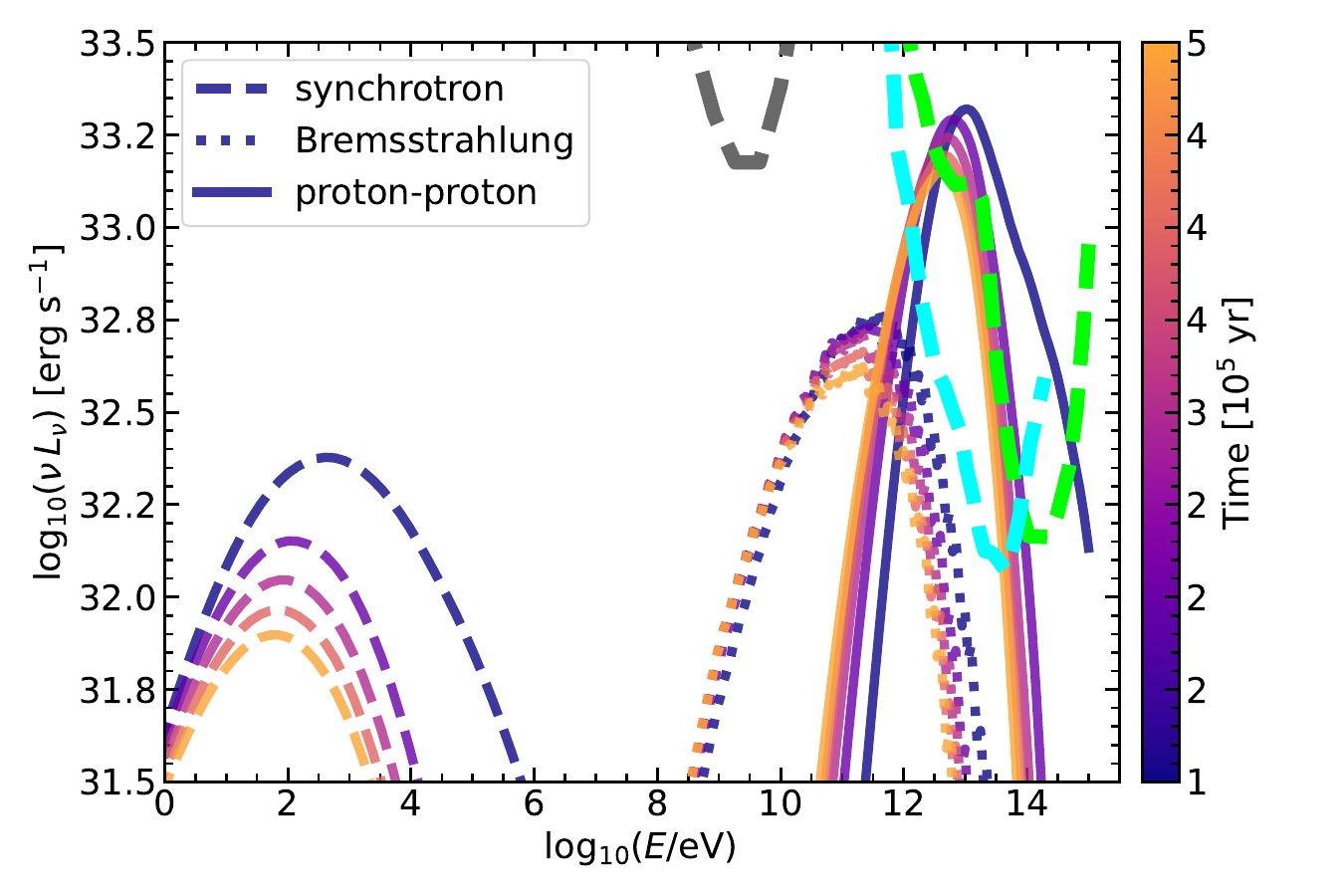}
  \end{minipage}
  \vspace{-0.4cm} % <-- reduce separación vertical

  \caption{\footnotesize SEDs of the illuminated clouds at different times (see color bar). Leptonic emission is produced by secondary pairs created in the cloud via $pp$ interactions. Thick dashed lines show the sensitivity curves of \textit{Fermi} (10 yr, grey), SWGO (5 yr, light blue), and LHAASO (1 yr, green) for a Galactic MQR at 5 kpc. \textit{Top}: Cloud in the MQR. \textit{Bottom}: Cloud at 100 pc from the MQR (continuous scenario).}
  \label{fig: seds}
\end{figure}

Long after the MQ activity ceases, the relativistic particles accumulated in the cocoon begin to leak into the ISM. These escaping CRs diffuse outward and can interact with nearby molecular material, leading to delayed gamma-ray emission. \hyperref[fig: proton_distribution]{Figure~\ref{fig: proton_distribution}} (Bottom) shows the proton distribution at a fixed distance of $100\,$pc from the MQR for different arrival times of the CRs.  %This section describes this propagation process and estimates the resulting illumination of clouds surrounding the MQR. 

\hyperref[fig: seds]{Figure~\ref{fig: seds}} shows the spectral energy distributions (SEDs) of irradiated clouds inside the MQR (Top) and 100 pc away from the MQR (Bottom). In both cases, we calculated the emission at different times (see the color bar). It can be seen that, even for $t\gg t_0$, UHE gamma rays are produced in the CR-cloud interaction with enough power to be detected by LHAASO for a source at 5 kpc from Earth %\textbf{We provide in the Appendix alternative results using non-Bohmian diffusion coefficients.} 
(see Appendix \ref{app: ISM particles} for details on particle transport in the ISM and clouds irradiation).

\section{Discussion and conclusion}\label{sect: conclusion}

Irradiated clouds would appear as VHE gamma-ray sources, while the PeVatron accelerating the CRs could remain undetected. At 5 kpc, the MQR would span $\sim 2^{\circ}$, but its radio surface brightness, produced by the primary electrons, would be below that of a bright supernova remnant, making it invisible to interferometers and hard to detect for single-dish surveys unless the large-scale Galactic diffuse emission is efficiently subtracted (e.g. \citealt{Combi1998A&A}). We estimate a moderate radio surface brightness of $\Sigma_{1.4 \,\textrm{GHz}}\sim 10^{-21} \; \textrm{W} \,\textrm{m}^{-2} \,\textrm{Hz}^{-1} \,\textrm{sr}^{-1}
$ for most of the MQR phase, assuming a magnetic field of $2.5\mu\,$G in the cocoon. This low surface brightness arises from two complementary effects. First, only about one percent of the non-thermal power is injected into electrons, which is consistent with the adopted proton-to-electron energy ratio (Sect. \ref{sect: relativistic particles}). Second, the source is very extended, and therefore the emissivity per unit area is very low. %Second, relativistic electrons in the cocoon undergo synchrotron and inverse-Compton cooling due to the local magnetic and radiation fields, which rapidly deplete their high-energy population.
%Electrons accelerated at the termination shock diffuse throughout the cocoon by the same turbulent flows that transport protons. However, they cool efficiently during this process and do not accumulate significant energy density. 
This explains why most MQ bubbles exhibit only faint extended radio emission despite their large kinetic powers. %Thus, the low radio brightness predicted here naturally follows from the combined effects of limited electron energy input and the spatial extension of the source.

%\textbf{Electrons accelerated at the reverse shock suffer strong synchrotron and inverse-Compton cooling; their energy distribution rapidly steepens above the GeV–TeV range, leaving only a faint radio component, consistent with the expected radio-quiet nature of MQRs.}

A single MQR could also generate multiple gamma-ray sources if located in a dense and structured ISM, suggesting a strategy to search for weak, extended radio counterparts in regions of clustered VHE emission.
Clouds at different distances from the CR reservoir show spectra modified by energy-dependent diffusion, with steepening at larger distances \citep{1996A&A...309..917A}. This behavior (see Eq.~\ref{eq:diffISM}) can be used to constrain the ISM diffusion coefficient.

The MQR lifetime is set by its ability to confine CRs, since the escape timescale grows as diffusion decreases. Cocoons approaching the Bohm limit ($\zeta=5$ here, but smaller values yield longer lifetimes) act as long-lived CR reservoirs. Even after the MQR dissipates, low diffusion in the environment can still generate a sequence of UHE, VHE, and HE gamma-ray sources at different distances from the remnant if a cloudy medium is present. 

Our conclusions rely on the assumption of quasi-Bohm diffusion inside the cocoon. If diffusion is faster, the CRs escape more rapidly and the $\gamma$-ray emission fades on shorter timescales. To test the robustness of this assumption, we explored alternative turbulence regimes (Kolmogorov and Kraichnan; see Appendix \ref{app}). In these cases, the hard power-law energy dependence of the diffusion coefficient reduces particle confinement, so that the MQR acts as a PeVatron for a shorter period than in the Bohm scenario. The resulting particle spectra lead to weaker non-thermal emission from the clouds in the cocoon and in the ISM: the $\gamma$-ray output shifts to lower energies, and the UHE component becomes strongly suppressed. These effects depend on the poorly constrained turbulence spectrum of MQ cocoons, but could be partially mitigated if higher magnetic fields or denser, nearby clouds are assumed.

The MQ type considered here is a super-Eddington system with powerful jets and hot spots for particle acceleration, exemplified by S26 in NGC 7793 \citep{Abaroa_etal2024(S26)}, and resembling FRII radio galaxies. A second class of MQs, with comparable power but slower jets ($v_{\rm j}\sim 0.3\,c$), resembles FRI galaxies and lacks bright terminal regions. SS 433 is a well-known case, detected by LHAASO at $>200$ TeV. In such systems, particle acceleration likely occurs at recollimation shocks, where lateral jet expansion is confined by external pressure; in SS 433, this happens at $\sim 30$ pc from the central engine \citep{SS433-2018Natur.562...82A,SS433-2024Sci}. The scenario discussed here applies to these sources with only minor modifications, and the overall conclusions remain valid.

In summary, we suggest that undetected remnants of extinct super-Eddington MQs may act as CR reservoirs illuminating nearby clouds, potentially explaining many unidentified LHAASO sources in the Galactic plane. Given the presence of active super-Eddington MQs in the Galaxy, such remnants are likely, and we have outlined their possible manifestations.

\begin{acknowledgements}
We thank the referee for his/her insightful comments, which helped to improve this work. LA thanks the Universidad Nacional de La Plata. GER and VBR were funded by PID2022-136828NB-C41/AEI/10.13039/501100011033/ and through the ``Unit of Excellence María de Maeztu'' award to the Institute of Cosmos Sciences (CEX2019-000918-M, CEX2024-001451-M). VB-R is Correspondent Researcher of CONICET, Argentina, at the IAR.
\end{acknowledgements}

\bibliographystyle{aa} % style aa.bst
\bibliography{main} % your references Yourfile.bib

\begin{appendix}

\section{System parameterization}

\subsection{General parameters} \label{app: table}

In the Table \ref{tab: parametros generales} below, we provide the main parameters of our model for the jet, cocoon, and clouds.

\begin{table}[h!]
\begin{center}
\caption{Parameters of the model.}
\label{tab: parametros generales}
\begin{adjustbox}{max width=\columnwidth}
\begin{tabular}{l c c c}
\hline
\hline
\rule{0pt}{2.5ex}Parameter & Value & Units  \\
\hline
\rule{0pt}{2.5ex}Age of the MQ [$t_0$]  & $10^5$ & ${\rm yr}$ \\
Jet mechanical power [$L_{\rm j}$]   & $10^{40}$ & ${\rm erg\,{s}^{-1}}$\\
%Jet length at $t_0$ $^{(2)}$[$l_{\rm j}$]   & $3\times10^{20}$ & ${\rm cm}$\\
Jet Lorentz factor [$\gamma_{\rm j}$]   & $3$ & \\
Magnetic field at RS   & $10$ & $\mu{\rm G}$  \\
Shock velocity [$v_{\rm sh}$]   & $2.8\times10^{10}$ & ${\rm cm\,s^{-1}}$  \\
%Number density of protons at RS$^{(1)}$[$n_{\rm l}$]   & 0.1  & ${\rm cm^{-3}}$\\
%Cocoon volume at $t_0$$^{(2)}$[$V_{\rm c}$]  & $10^{61}$ & ${\rm cm^3}$\\
%Cocoon pressure$^{(2)}$[$p_{\rm c}$] & $4.5\times10^{-11}$ & ${\rm erg\, cm^{-3}}$\\
Cocoon number density [$n_{\rm cocoon}$]  & $10^{-3}$ & ${\rm cm^{-3}}$\\
ISM number density [$n_{\rm ISM}$]  & $0.1$ & ${\rm cm^{-3}}$\\
Number density (cloud inside cocoon)  & $10^3$ & ${\rm cm^{-3}}$\\
Number density (cloud in ISM)  & $10^5$ & ${\rm cm^{-3}}$\\
Cloud radius (both clouds) [$R_{\rm cl}$]  & $1$ & ${\rm pc}$\\
Cloud magnetic field (both clouds) [$B_{\rm cl}$]  & $10^{-4}$ & ${\rm G}$\\
%Cloud volume (in MQR) $^{(2)}$[$V_{\rm cl}$] & $3.8\times10^{56}$ & ${\rm cm^3}$\\
%Cloud surface density $^{(1)}$[$l_{\rm c}$]  & $27$ & ${\rm cm^{-2}}$\\
\hline

\end{tabular}
\end{adjustbox}
\end{center}
\footnotesize{\textbf{Notes.} %All parameters are assumed.
Clouds' parameters from \cite{Bergin&Tafalla2007} and \cite{crutcher2012}.}
\end{table}

\subsection{Cocoon survival}\label{app: cocoon survival}

Here, we describe the characterization of the cocoon as it expands into the ISM, and the conditions for survival. The cocoon begins to break and to mix with the ISM in the transition between the adiabatic and radiative phases.

The transition occurs when the cooling time $t_{\rm cool}(t)=3k_{\rm B}T(t)/n\Lambda(T(t))$ becomes equal to $t$ \citep{castor_etal_1977}. Here $k_{\rm B}$ is the Boltzmann constant, $\Lambda$ the cooling function, and $n$ is the electron density (we assume a plasma of pure hydrogen, so that the mean molecular weight in the ionized medium is $\mu=0.5$). The temperature in the cocoon is given by $T_{\rm c}(t)=2E_{\rm int}(t)/3n_{\rm cocoon}V_{\rm c}(t)k_{\rm B}$, where $E_{\rm int}(t)\sim 0.5L_{\rm j}t$; at $t=t_0$, we obtain $T_{\rm c}\approx 6\times 10^8\,$K. As expected for jet-inflated bubbles, this implies that roughly half of the jet's kinetic energy is converted into the cocoon’s internal energy, rather than representing an independent energy channel, as is the case in purely wind-driven outflows \citep{Peretti2025}. The temperature for the expanding shell is $T_{\rm s}(t)=3\mu m_{\rm p}v_{\rm c}^2/16k_{\rm B}$ (where %$\mu$ is the mean molecular weight and 
$m_{\rm p}$ is the proton mass): %$t_{\rm rad}\gg t_0$.
at $t=t_0$, we get $T_{\rm s}\approx 3.3\times10^6\,{\rm K}$. Considering a post-shock density of $0.4\,{\rm cm^{-3}}$ and a cooling function value of $\Lambda(T)\sim 2\times 10^{-22}\,{\rm erg\, cm^{3}\,s^{-1}}$ for the temperature $T_{\rm s}$ above, we obtain $t_{\rm cool}\approx 5\times 10^{5}\,{\rm yr}$ for the shell, i.e., $t_{\rm rad}\approx 5t_0.$

%Inside the cocoon the diffusion properties are different from those of the ISM. The internal pressure is given by $p_{\rm c}=(3/4) \rho_{\rm ISM}v_{\rm c}^2$, while the ISM pressure is $p_{\rm ISM}=n_{\rm ISM}k_{\rm B}T_{\rm ISM}\approx 10^{-16}\,{\rm erg\,{cm^{-3}}}$ (where $T_{\rm ISM}=10\,{\rm K}$ corresponds to the temperature of a star-forming region, see e.g. \citealt{Schmalzl2014}). Comparing both quantities, we find that the cocoon is overpressured for  $t<\mathbb{C}\, \rho_{\rm ISM}^{-1/2}\,L_{\rm j}^{1/2}\,T_{\rm ISM}^{-5/4}$ (with $\mathbb{C}=4.4\times 10^{-8}\, {\rm g^{1/2}\, erg^{1/2}\, s^{3/2}\, K^{5/4}}$), a condition always satisfied for the parameters assumed in our model. %The internal pressure is given by $p_{\rm c}=(3/4) \rho_{\rm ISM}v_{\rm l}^2$; we obtain $3\times10^{-12}\,{\rm erg\,cm^{-3}}$ ($A_1$) and $4.5\times10^{-11}\,{\rm erg\,cm^{-3}}$ ($A_2$) for $t=t_0$. On the other hand, the ISM pressure is $p_{\rm ISM}=n_{\rm ISM}k_{\rm B}T_{\rm ISM}\approx 10^{-16}\,{\rm erg\,{cm^{-3}}}$ (where $T_{\rm ISM}=10\,{\rm K}$ corresponds to the temperature of a star-forming region, see e.g. \citealt{Schmalzl2014}). Comparing both pressures, we find that the cocoon is overpressured for  $t<\mathbb{C}\, \rho_{\rm ISM}^{-1/2}\,L_{\rm j}^{1/2}\,T_{\rm ISM}^{-5/4}$ (with $\mathbb{C}=4.4\times 10^{-8}\, {\rm g^{1/2}\, erg^{1/2}\, s^{-1/2}\, K^{5/4}}$) a condition always satisfied for the parameters assumed in our model. 

\subsection{Clouds engulfment}

The cocoon expansion can lead to the engulfment of nearby clouds (see \hyperref[tab: parametros generales]{Table~\ref{tab: parametros generales}}  for cloud parameters). We follow the criteria of \cite{klein_etal_1994} to assess whether a cloud engulfed by the expanding cocoon would survive.

We compare the timescales associated with crushing due to Kelvin-Helmholtz or Rayleigh-Taylor instabilities ($t_{\rm cc}=\sqrt{\chi}R_{\rm cl}/v_{\rm c}\approx t_{\rm KH} \sim t_{\rm RT}$, where $\chi=n_{\rm cl}/n_{\rm ISM}=10^4$, $v_{\rm c}$ is the expansion velocity of the cocoon, and $R_{\rm cl}$ the cloud radius), shredding ($t_{\rm sh}=Nt_{\rm cc}$, with $N>1$ a dimensionless factor that depends on the cloud's magnetic field), and dragging $(t_{\rm drag}=\chi t_{\rm cc})$. %, and exposure to the expansive cocoon $(t_{\rm exp}=R_{\rm cl}/v_{\rm c})$.
On the other hand, the magnetic Mach number is given by $M_{\rm A}=v_{\rm sc}/v_{\rm A}$, where $v_{\rm sc}=v_{\rm c}/\sqrt{\chi}$ is the shock velocity in the cloud and $v_{\rm A}=B_{\rm cl}/\sqrt{4\pi \rho_{\rm cl}}$ is the Alfvén velocity. If $M_{\rm A}<1$, the shock is sub-Alfvénic, the magnetic field drapes the cloud and suppresses hydrodynamic instabilities in the direction perpendicular to the local field lines \citep{McCourt_etal_2015MNRAS}. The post-shock temperature is given by $T_{\rm ps}\sim 10^5 (v_{\rm sc}/100\,{\rm km\,s^{-1}})^2\,{\rm K}$. %If we suppose that the cloud is engulfed at $t=t_0$, we have: $v_{\rm c}\sim 542\,{\rm km\,s^{-1}}$, $t_{\rm cc}\sim 5.5\times 10^{12}\, {\rm s}$, $t_{\rm sh}\sim 5.5\times 10^{13}\, {\rm s}$, $t_{\rm drag}\sim 5.5\times 10^{16}\, {\rm s}$, $v_{\rm sc}\sim 5.4\,{\rm km\, s^{-1}}$, $v_{\rm A}\sim 9\,{\rm km\, s^{-1}}$, $M_{\rm A}\approx0.63$, $t_{\rm exp}=5.5\times 10^{10}\, {\rm s}$, and $T_{\rm ps}\approx 291\,{\rm K}$. Therefore, given the parameters of our model, the cloud survives the engulfment $(t_{\rm sh}\gg t_{\rm exp})$, it is not dragged $(t_{\rm drag}\gg t_{\rm exp})$, and we can neglect the cloud ionization $(T_{\rm ps}\ll 10^4\,{\rm K})$.
If we suppose that the cloud is engulfed at $t=t_0$, we have: $v_{\rm c}\approx 542\,{\rm km\,s^{-1}}$, $t_{\rm cc}\approx 5.5\times 10^{12}\, {\rm s}\, \sim 2\,t_0$, $t_{\rm sh}\approx 5.5\times 10^{13}\, {\rm s}$, $t_{\rm drag}\approx 5.5\times 10^{16}\, {\rm s}$, $v_{\rm sc}\approx 5.4\,{\rm km\, s^{-1}}$, $v_{\rm A}\approx 9\,{\rm km\, s^{-1}}$, $M_{\rm A}\approx0.6$, and $T_{\rm ps}\approx 291\,{\rm K}$. Therefore, the cloud survives the engulfment $(t_{\rm sh}\gg t_{\rm cc})$, it is not dragged $(t_{\rm drag}\gg t_{\rm cc})$, and we can neglect the cloud ionization $(T_{\rm ps}\ll 10^4\,{\rm K})$.

\section{Particle transport}

\subsection{Transport in cocoon}\label{app: cocoon transport}
We inject the relativistic particles from the RS in the cocoon and determine their evolution by solving the time-dependent transport equation in the one-zone approximation:
\begin{equation}
   \frac{{\partial n_{\rm c}(E,t)}}{\partial t}+ \frac{\partial}{\partial E}\left[\frac{{\rm d}E}{{\rm d}t} n_{\rm c}(E,t)\right]+\frac{n_{\rm c}(E,t)}{t_{\rm esc, c}(E,t)}=Q_{\rm c}(E,t).
\end{equation}
Here, the injection is given by $Q_{\rm c}(E,t)=n_{\rm RS}(E,t)V_{RS}(t)/t_{\rm esc, RS}(t) V_{\rm c}(t)$, where $V_{\rm RS}\sim \Delta x^3$ and $V_{\rm c}$ are the volumes of each zone, and $n_{\rm RS}$ is the proton distribution in the RS. Escape of the cocoon is diffusive ($t_{\rm esc,c}\equiv t_{\rm diff}$).

\subsection{Transport in ISM}\label{app: ISM particles}

Equation \eqref{eq:diffISM} below expresses the transport of CRs in the ISM once they have escaped the cocoon, following the formalism of \cite{1996A&A...309..917A}. %The CRs that escape from the MQR are injected into the ISM, where they propagate by diffusion. 
Considering propagation in the spherically symmetric case, the transport equation of protons reads:
\begin{dmath}
     \dfrac{\partial n_{\rm I}(E,R,t)}{\partial t}=\dfrac{D(E)}{R^2}\dfrac{\partial}{\partial R}\left[R^2 \dfrac{\partial n_{\rm I}(E,R,t)}{\partial R}\right]++\dfrac{\partial}{\partial E}\left[\left(-\dfrac{{\rm d}E}{{\rm d}t}\right)\,n_{\rm I}(E,R,t)\right]+Q(E,R,t), \label{eq:diffISM}
    \end{dmath}
where the injection function is $Q(E,R,t)=Q_0(E,t)\delta(R)$. Here, $Q_0(E,t)=n_{\rm c}(E,t)V_{\rm c}(t)/t_{\rm esc,c}(E,t)$ and $\delta(R)$ is a delta function in space. A solution to this equation is provided by \cite{1996A&A...309..917A}. We consider the scenario of a continuous injection of energy. 
%impulsive injection of energy at $t=t_0$.

The diffusion coefficient in the ISM is given by $D(E)=D_0 E^{\delta}$, where we assume that $\delta=0.5$ and that the diffusion coefficient normalization constant is $D_0=D(10\,{\rm GeV})=10^{26}\,{\rm cm^2\,s^{-1}}$ (e.g. \citealt{1996A&A...309..917A}). This value is a suppressed version of the average Galactic coefficient ($D=\chi D_{\rm gal}$, with $\chi=10^{-2}$ and $D_{\rm gal}(10\,{\rm GeV})\approx 10^{28} \,{\rm cm\, s^{-1}}$), as the environment surrounding a former MQ is expected to be more turbulent and magnetically irregular.
Such conditions may locally suppress diffusion because of the Alfvén-wave excitation by the escaping CRs themselves.
This leads to a slow-diffusion zone analogous to those inferred around other compact accelerators \citep{1996A&A...309..917A}. This assumption allows the CRs to remain concentrated near the source and improves their probability of interaction with nearby clouds. %\textbf{We provide in the Appendix alternative results using non-Bohmian diffusion coefficients.}

Proton-proton interactions between CRs and clouds lead to the production of neutral pions, which decay into gamma rays ($\pi^0 \rightarrow 2\gamma$), and charged pions, which in turn decay into electron-positron pairs ($ \pi^{\pm}\rightarrow \mu^{\pm}+\nu_{\mu}\rightarrow e^{\pm}+ {\rm neutrinos}$). These secondary pairs then interact with matter and magnetic fields, producing synchrotron and Bremsstrahlung radiation. Inverse Compton radiation can be neglected (see \citealt{2005A&A...432..609B}).

%The fraction of CRs that penetrate a cloud depends on several factors, including the magnetic field, the diffusion coefficient, advection, and energy losses due to ionization of neutral hydrogen atoms \citep{Cesarsky1978A&A,Morlino2015MNRAS,Dogiel2018ApJ,Ivlev2018ApJ, Abaroa_etal2024(S26)}. 

\section{Non-Bohmian diffusion}\label{app}

To test the sensitivity of our results to the assumed turbulence regime, we examined diffusion coefficients following Kolmogorov and Kraichnan scalings. The particle diffusion coefficient can be written as \citep{Ptuskin2012,Peretti2025}

\begin{equation}
D(E) \approx \frac{1}{3}\,c\,L_{\mathrm{c}}
\left( \frac{r_{\mathrm{L}}(E)}{L_{\mathrm{c}}} \right)^{\!\delta}
\left( \frac{B}{\delta B} \right)^{\!2},
\quad
\delta =
\begin{cases}
1   & \text{(Bohm)},\\[4pt]
1/2 & \text{(Kraichnan)},\\[4pt]
1/3 & \text{(Kolmogorov)}.
\end{cases}
\label{eq:diffusion_general}
\end{equation}

Here, $r_{\mathrm{L}}(E) = E/eB$ is the Larmor radius of a relativistic particle with energy $E$, $B$ is the mean magnetic field, $\delta B$ its turbulent component, and $L_{\mathrm{c}}$ the coherence (or injection) scale of turbulence, where most of the magnetic energy is contained. In the Bohm limit $(\delta = 1)$ the diffusion coefficient reaches its minimum value.

For non-Bohmian regimes $(\delta = 1/2,\,1/3)$, we adopt $L_{\mathrm{c}} \sim 10^{-2}\,l_{\rm c}$ at each time $t$, corresponding to the expected correlation length of turbulence within the cocoon (a small but reasonable fraction of its radius; see \citealt{Zhang2025arXiv250620193Z}). This parameterization enables comparison among different turbulence spectra, while assuming that the acceleration process—occurring at the relativistic jet termination shock—operates near the Bohm regime, which sets the maximum particle energy $E_{\max}$ in our model.

\begin{figure}[ht!]
  \centering
  \begin{minipage}{0.5\textwidth}
    \centering
    \includegraphics[width=8.3cm, height=5cm]
    {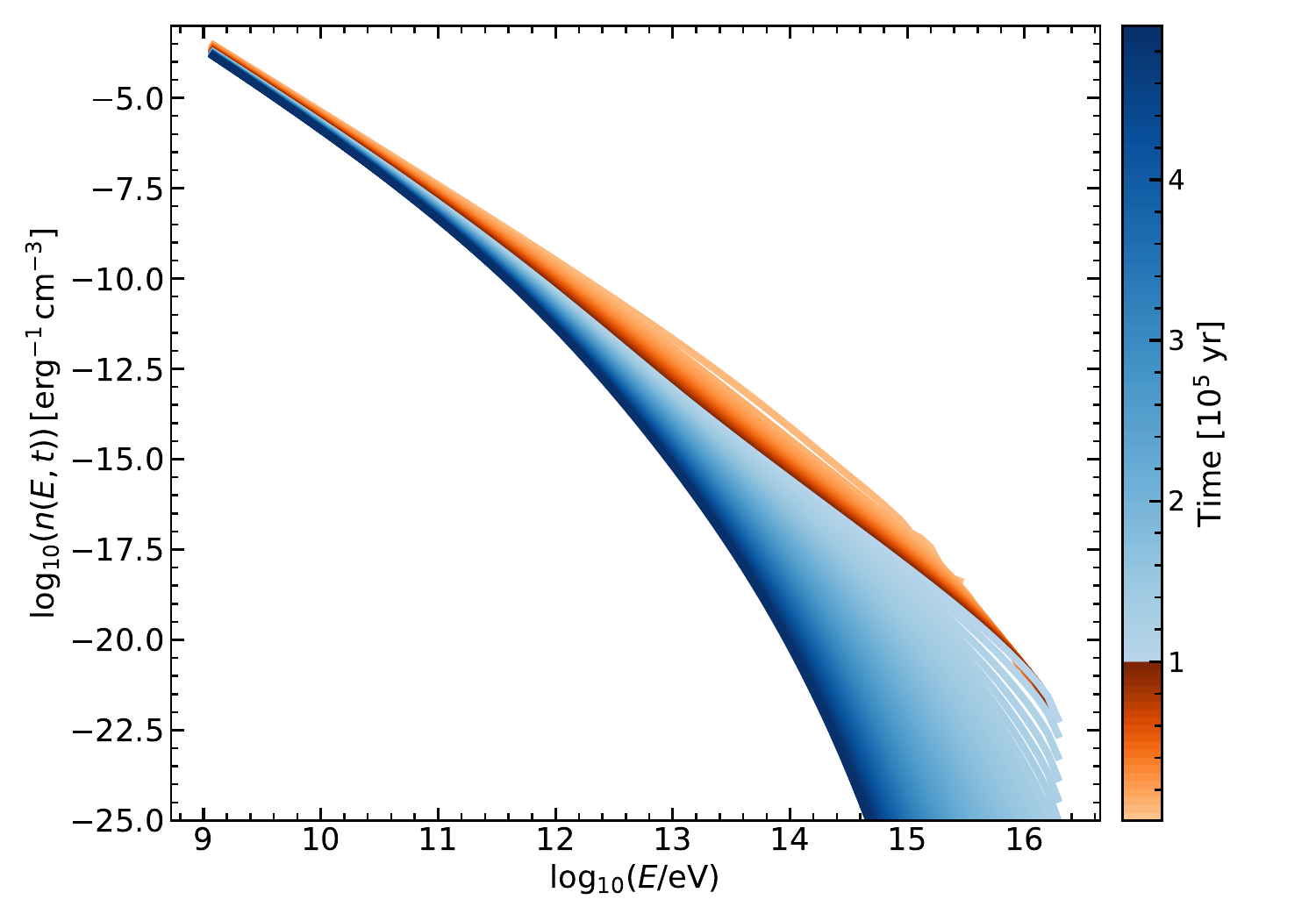}
  \end{minipage}

  \vspace{-0.2cm} % <-- reduce separación vertical

  \begin{minipage}{0.5\textwidth}
    \centering
    \includegraphics[width=7.7cm, height=5cm]{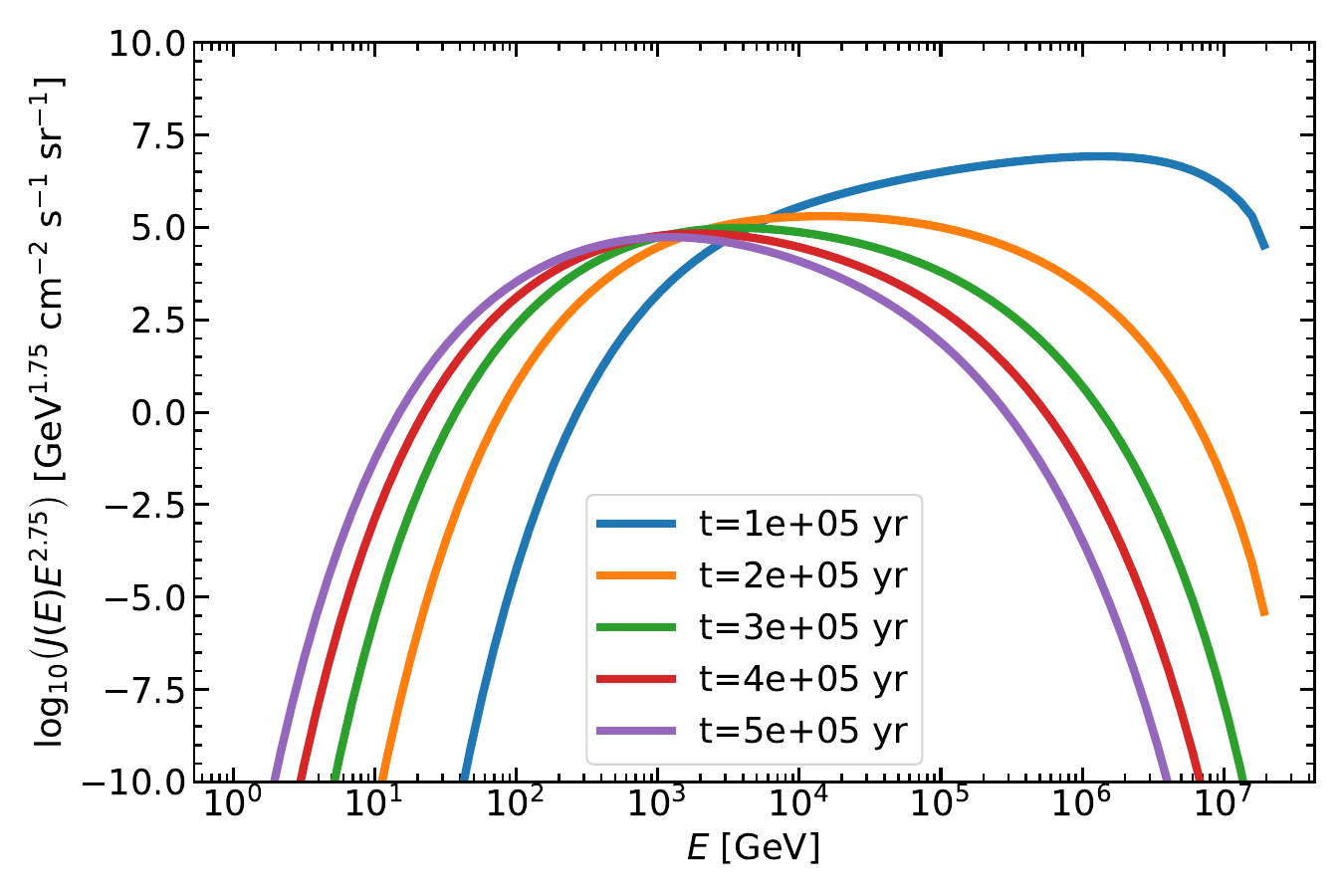}
  \end{minipage}
  \vspace{-0.4cm} % <-- reduce separación vertical

  \caption{\footnotesize Same as in Fig.\ref{fig: proton_distribution} but assuming Kolmogorov diffusion inside the cocoon.}
    \label{fig: proton_distribution_kolm}
\end{figure}

\begin{figure}[ht!]
  \centering
  \begin{minipage}{0.5\textwidth}
    \centering
    \includegraphics[width=8.3cm, height=5cm]{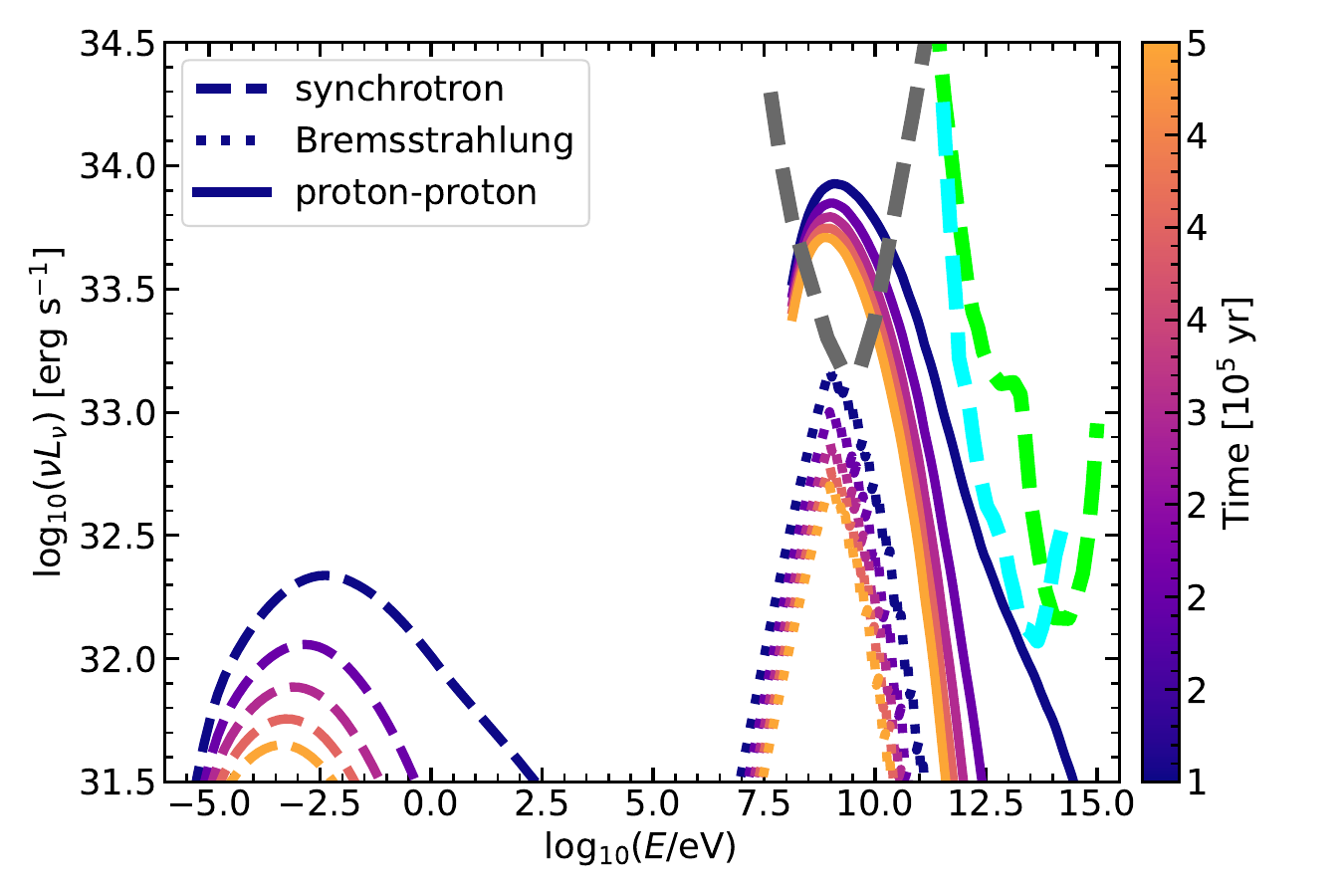}
  \end{minipage}

  \vspace{-0.2cm} % <-- reduce separación vertical

  \begin{minipage}{0.5\textwidth}
    \centering
    \includegraphics[width=8.3cm, height=5cm]{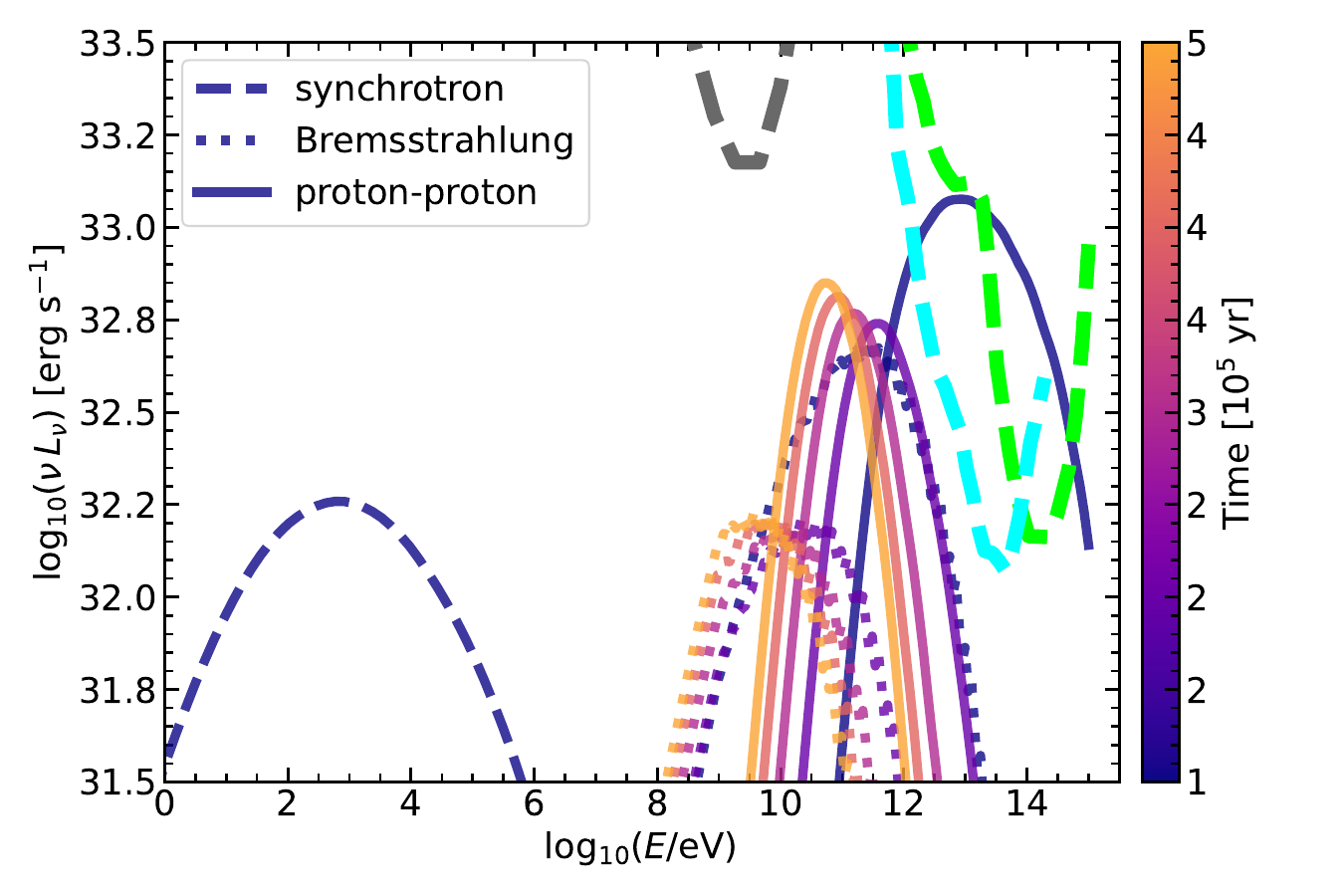}
  \end{minipage}
  \vspace{-0.4cm} % <-- reduce separación vertical

  \caption{\footnotesize Same as in Fig. \ref{fig: seds} but assuming Kolmogorov diffusion inside the cocoon.}
  \label{fig: seds_kolm}
\end{figure}

As an illustrative example, Figs.~\ref{fig: proton_distribution_kolm} and~\ref{fig: seds_kolm} show the proton distributions (in the cocoon and ISM) and the corresponding SEDs (for the MQR and ISM clouds) for Kolmogorov diffusion with $\delta B/B = 0.3$. In this case, the MQR cloud emits detectable $\gamma$-rays up to $\sim 10\,$TeV for $\lesssim 10^5\,$yr after the MQ shuts down, although the emission at UHEs is markedly suppressed. The ISM cloud also shows a noticeable change: its $\gamma$-ray spectrum shifts to slightly lower energies and the overall luminosity decreases compared to the Bohm case, although the effect is less pronounced than in the MQR cloud. We also tested the Kraichnan regime and different turbulence levels (not shown here). The qualitative behavior remains similar, although Kraichnan diffusion is slightly more favorable to UHE emission.

%\textbf{We also tested the Kraichnan regime and different turbulence levels ($\delta B/B$). The qualitative behavior remains similar, although Kraichnan diffusion is slightly more favorable to UHE emission. Under Kolmogorov diffusion, the $\gamma$-ray luminosity above 100\,TeV decreases by roughly an order of magnitude compared to the Bohm scenario.}

\end{appendix}

\end{document}